\documentclass[aps, superscriptaddress, twocolumn]{revtex4-1}
\usepackage{hyperref}
\usepackage{graphicx}
\usepackage{amsmath}
\usepackage{amssymb}
\usepackage{amsfonts} 
\usepackage{braket}
\usepackage{color}
\usepackage{bm}

\newcommand{\Tr}{\text{Tr}}

\newcommand{\up}{\uparrow}
\newcommand{\dn}{\downarrow}

\begin{document}
\title{Effect of mediated interactions on a Hubbard chain in mixed-dimensional fermionic cold atoms}

\author{Junichi Okamoto}
\affiliation{Physikalisches Institut, Albert-Ludwigs-Universit\"at Freiburg, Hermann-Herder-Stra\ss e 3, 79104 Freiburg, Germany}

\author{Wen-Min Huang}
\email{wenmin@phys.nchu.edu.tw}
\affiliation{Department of Physics, National Chung-Hsing University, Taichung 40227, Taiwan}

\author{Kyle Irwin}
\affiliation{Department of Physics and Astronomy, University of California, Riverside, CA 92521, USA}

\author{David K. Campbell}
\affiliation{Department of Physics, Boston University, Boston, Massachusetts 02215, USA}

\author{Shan-Wen Tsai}
\affiliation{Department of Physics and Astronomy, University of California, Riverside, CA 92521, USA}

\date{\today}

\begin{abstract}
Cold atom experiments can now realize mixtures where different components move in different spatial dimensions. We investigate a fermion mixture where one species is constrained to move along a one-dimensional lattice embedded in a two-dimensional lattice populated by another species of fermions, and where all bare interactions are contact interactions. By focusing on the one-dimensional fermions, we map this problem onto a model of fermions with non-local interactions on a chain. The effective interaction is mediated by the two-dimensional fermions and is both attractive and retarded, the form of which can be varied by changing the density of the two-dimensional fermions. By using the functional renormalization group in the weak-coupling and adiabatic limit, we show that the one-dimensional fermions can be controlled to be in various density-wave, or  spin-singlet or triplet superconducting phases.
\end{abstract}

\maketitle
\section{Introduction}\label{sec:intro}
Cold atom systems have proven to be an invaluable tool in studying a wide range of quantum many-body phenomena. Even a single species of atoms can exhibit various quantum states such as Bose-Einstein condensates  (BECs)\cite{anderson1995, davis1995}, Mott insulators \cite{greiner2002}, or unitary Fermi gases \cite{bartenstein2004, regal2004, zwierlein2004, bourdel2004}. The complexity of cold atomic systems can be greatly enhanced by mixing different atoms. For example, a degenerate Fermi gas immersed in a BEC can be realized 
with isotopes of Li atoms \cite{schreck2001, truscott2001, ferrier-barbut2014} or 
with different atomic species \cite{hadzibabic2002, gunter2006}. A two-species Fermi gas was first realized 
with Li and K atoms \cite{wille2008}, and has been studied extensively \cite{voigt2009,trenkwalder2011, jag2014}. The controllability of inter-species interactions permits these ultracold atomic mixtures to become valuable platforms to study novel many-body phenomena such as an impurity problem \cite{jaksch2005,kato2012}, polaron formation \cite{cucchietti2006,blume2012}, or lattice gauge theories \cite{zohar2016,zhang2018}.

The development of experimental techniques to create species-specific lattice geometries has further extended the possibilities of the mixed atomic gases \cite{leblanc2007, catani2009}. For example, in a recent experimental work, a bosonic mixture of $^{41}$K and $^{87}$Rb atoms is confined in two dimensions and in three dimensions, respectively \cite{lamporesi2010}. In the light of this experimental success, several other experimental works \cite{haller2010, schafer2018} have been conducted to create Fermi-Bose mixtures. We also note that a $^3$He-$^4$He solution provides effectively mixed-dimensional Fermi gases \cite{ikegami2017}. Theoretical works on mixed-dimensional systems predict phenomena such as Efimov effects \cite{nishida2008, nishida2010}, mediated-pairing \cite{iskin2010, huang2013, okamoto2017a,caracanhas2017, kelly2018} or a topological superfluid \cite{wu2016, midtgaard2016}. 

\begin{figure}[!tbp]
\begin{center}
   \includegraphics[width = 0.65\columnwidth ]{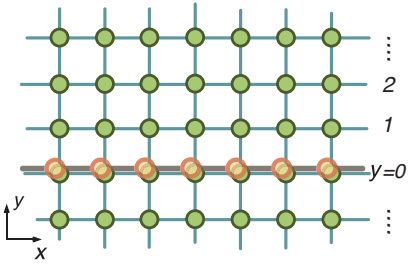}
   \caption{Schematic of the model that we consider. The one-dimensional optical lattice is in contact with the two-dimensional lattice at $y=0$.}
\label{fig:model}
\end{center}
\end{figure}

Motivated by this progress, we propose in the present work to investigate a Fermi system composed of $c$-type and $f$-type species confined to one dimension and two dimensions, respectively (Fig.~\ref{fig:model}). We assume a contact interaction in the one-dimensional (1D) system while ignoring the interaction between the two-dimensional (2D) particles. Through the inter-species contact interaction, the $f$-particles induce mediated interaction of the Ruderman-Kittel-Kasuya-Yosida-type \cite{ruderman1954, kasuya1956, yosida1957} among the $c$-particles. The momentum structure of the effective interaction is determined by the filling of the $f$-particles and thus can be used as a tuning knob to manipulate the quantum phases in the 1D system.

In the weak-coupling limit, we investigate the ground state phase diagram by a functional renormalization group (fRG) \cite{shankar1994, halboth2000, tam2006,tam2014, kopietz2010, metzner2012, platt2013}. This treatment enables us to go beyond the standard g-ology and bosonization methods \cite{voit1995} by including the curvature of the band dispersion and the momentum-dependence of the coupling constants. We identify the dominant instability by looking at the flow of the coupling constants and show that the resulting phase diagram includes charge/bond/spin-density-waves and spin-singlet/triplet superconductivity. We find that formally irrelevant terms can play an important role in determining the dominant pairing phase, and also lead to a dominant bond-wave order instability among the many competing ones in a small region of the phase space. When the filling of the $c$-particles is away from half-filling, the density-wave instabilities are replaced by the superconducting instabilities. This demonstrates a possible means of manipulating exotic quantum phases such as bond-density waves or spin-triplet pairing in a mixed-dimensional setup.

The remainder of our paper is organized as follows. Section \ref{sec:model} introduces the model that we consider and discusses the mediated interactions in the 1D system. In Sec.~\ref{sec:method}, we briefly summarize the functional renormalization group method as applied to our model, and explain how we determine the phase diagram. Section \ref{sec:results} shows the flows of the coupling constants and the phase diagrams. Section V presents our conclusions.

\section{Model}\label{sec:model}
The 1D system that we consider is given by
\begin{multline}
H_c = -t_c \sum_{\langle x x'\rangle, \sigma} \left( c^{\dagger}_{x, \sigma} c_{x', \sigma}  + \text{h.c.} \right) \\
+ U_c \sum_{x} n^c_{x,\up} n^c_{x,\dn} - \mu_c \sum_{x,\sigma}n^c_{x\sigma},
\end{multline}
where $c^{(\dagger)}_{x\sigma}$ is the annihilation (creation) operator of a $c$-fermion with spin $\sigma$ at site $x$, and $t_c$ is the hopping between nearest-neighbor sites $\langle x, x'\rangle$ on a regular lattice. $U_c$ is the on-site Hubbard interaction between particle densities $n^c_{x \sigma} = c^{\dagger}_{x\sigma}c_{x\sigma}$, and $\mu_c$ is the chemical potential. Similarly, the 2D system of the noninteracting $f$-fermions is given by
\begin{equation}
H_f = -t_f  \sum_{\langle \bm{r}, \bm{r}'  \rangle ,\sigma} f_{\bm{r} \sigma}^{\dagger} f_{\bm{r}' \sigma} - \mu_f \sum_{\bm{r}, \sigma} n^f_{\bm{r}, \sigma},
\end{equation}
where $f^{(\dagger)}_{\bm{r}\sigma}$ is the annihilation (creation) operator of a $f$-fermion with spin $\sigma$ at site $\bm{r} = (x,y)$, and $t_f$ is the hopping between nearest-neighbor sites $\langle \bm{r}, \bm{r}'\rangle$ on a square lattice. The two species interact at $y=0$ with spin-independent density-density interaction $U_{cf}$
\begin{equation}
H_{cf} = U_{cf} \sum_{x, \sigma, \sigma'} n^c_{x\sigma} n^f_{x,y=0,\sigma'}.
\end{equation} 
We consider $N_0$ and $N_0^2$ sites for $c$- and $f$-particles with 
periodic boundary condition with lattice constant $a=1$. The Fourier transforms are given by 
\begin{equation}
\begin{split}
c_{x \sigma} &= \frac{1}{\sqrt{N_0}} \sum_{k \in \text{BZ}} e^{i k x} c_{k \sigma}, \\ 
f_{\bm{r} \sigma} &= \frac{1}{\sqrt{N_0^2}} \sum_{\bm{p} \in \text{BZ}} e^{i \bm{p} \cdot \bm{r} } f_{\bm{p} \sigma},
\end{split}
\end{equation}
which give rise to the 1D and 2D dispersion relations $\xi^c_k = -2t_c \cos(k) - \mu_c$ and $\xi^f_{\bm{p}} = -2t_f \left[ \cos(p_x) + \cos(p_y) \right] - \mu_f$.

Next, we construct a path integral partition function over Grassmann fields [8], and seek an effective 1D action. The total action of the 1D and 2D fermions with imaginary time $\tau$ and inverse temperature $\beta$ is given by, 
\begin{multline}
S  = \int_0^{\beta} d\tau \Big[  \sum_{k, \sigma } c^{\dagger}_{k \sigma} (\tau ) \partial_{\tau}c _{k \sigma} (\tau )   + H_c(\tau )  \\
+ \sum_{\bm{p}, \sigma} f^{\dagger}_{\bm{p}\sigma} (\tau ) \partial_{\tau}f_{\bm{p}\sigma} (\tau ) + H_f(\tau )  + H_{cf}(\tau )  \Big].
\label{action}
\end{multline}
With the Fourier series in Matsubara frequencies $\omega_n = \pi (2n+1)/\beta$ ($n \in \mathbb{Z}$) as 
\begin{equation}
f_{\bm{p} \sigma} (\tau) = \frac{1}{\sqrt{\beta}} \sum_{n} e^{-i\omega_n \tau} f_{n \bm{p} \sigma},
\end{equation}
we can cast the quadratic action of the 2D fermions into a matrix form,
\begin{equation}
S_\text{2D} =\sum_{n, n', \bm{p},\bm{p}', \sigma}  f^{\dagger}_{n \bm{p} \sigma}  \left[- G_0^{-1} + M \right]_{(n \bm{p});(n' \bm{p}')} f_{n' \bm{p}' \sigma} ,
\end{equation}
with a Green's function matrix $G_0$ and an interaction part $M$,
\begin{equation}
\begin{split}
[G_0]_{(n \bm{p});(n' \bm{p}')} &= \frac{1}{ i \omega_n - \xi_f(\bm{p})} \delta_{nn'} \delta_{\bm{p} \bm{p}'} ,\\
[M]_{(n \bm{p});(n' \bm{p}')}  &= \frac{U_{cf}}{N_0^3 \beta }\sum_{m, k, \sigma}  c_{m k \sigma}^{\dagger} c_{n -n' +m, p_x - p'_x + k, \sigma}.
\end{split}
\end{equation}
After integrating out the 2D fermions, the effective action for the 1D fermions becomes
\begin{equation}
\begin{split}
S_\text{eff} &= S_c -  2\Tr \ln \left[ -G_0^{-1} + M \right] \\
&= S_c - 2\Tr \ln  \left[ -G_0^{-1}(1 - G_0 M) \right] \\
&= S_c + 2\sum_{n=1}^{\infty} \frac{ \Tr\left[ (G_0 M)^n \right]}{n} + \text{const},
\end{split}
\end{equation}
where the factor of $2$ accounts for the spin of the $f$-particles, taken here to be spin-1/2. The first-order term in the expansion gives the correction to the chemical potential of the 1D fermions, which we assume to be included in $\mu_c$ in the following. Summation over $\omega_n$ in the second-order term generates the effective interaction, 
\begin{multline}
\Tr\left[ (G_0 M)^2 \right] =  \frac{U_{cf}^2   }{N_0^6 \beta } \sum_{\tilde{l}, \bm{p},\bm{p}'} \Bigg[  \frac{n_\text{F}\left(\xi^f_{\bm{p}}\right) - n_\text{F}\left(\xi^f_{\bm{p}+\bm{p}'}\right)}{i \tilde{\omega}_l + \xi^f_{\bm{p}} - \xi^f_{\bm{p} + \bm{p}'}}     \\
\times \sum_{m,m', k,k', \sigma, \sigma'} c_{m k \sigma}^{\dagger} c_{m-\tilde{l}, -p_x'  + k, \sigma} c_{m' k' \sigma'}^{\dagger} c_{m' + \tilde{l}, p_x'  + k', \sigma'} \Bigg] ,
\label{eq:second-order}
\end{multline}
where $\tilde{\omega}_l = \omega_{n'} - \omega_n$ is a bosonic Matsubara frequency, and $n_\text{F}(\xi)$ is the Fermi distribution function. If we consider $t_f \gg t_c$, we can ignore the retardation effects, and only the $\tilde{\omega}_l = 0$ component is important. In this limit, we can write mediated interaction for the $c$-fermions, Eq.~\eqref{eq:second-order}, in momentum space as
\begin{equation}
\begin{split}
H_\text{med} &= \frac{1}{2N_0} \sum_{k, k', q, \sigma, \sigma'} U'(q) c^{\dagger}_{k \sigma}c^{\dagger}_{k' \sigma'}c_{k'+q \sigma'} c_{k-q \sigma}, \\
U'(q) &= 2 U_{cf}^2 \int_{-\pi}^{\pi} \frac{d\bm{p} dp'_y}{(2\pi)^3}   \frac{n_F\left(\xi^f_{\bm{p}} \right) - n_F\left(\xi^f_{\bm{p}+(q,p_y')}\right)}{ \xi^f_{\bm{p}} - \xi^f_{\bm{p}+(q,p_y')}},
\end{split}
\label{eq:Umed}
\end{equation}
where the strength of the mediated interaction scales as $U_{cf}^2/t_f$.

\begin{figure}[!tbp]
\begin{center}
   \includegraphics[width = 0.95\columnwidth ]{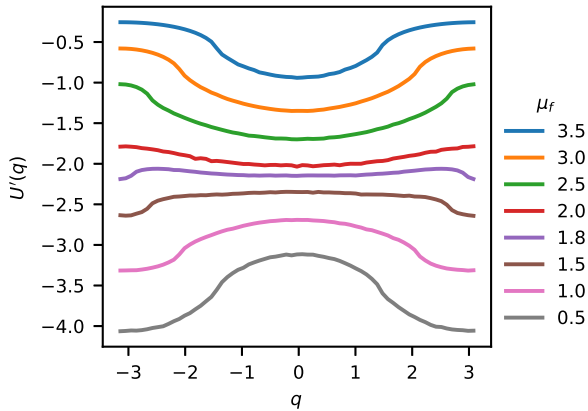}
   \caption{Mediated interaction $U'(q)$ in Eq.~\eqref{eq:Umed} in unit of $U_{cf}^2/t_f$. As the chemical potential of the $f$-particle increases, the dominant component shifts from $U'(\pi)$ to $U'(0)$.}
\label{fig:Uq}
\end{center}
\end{figure}

We plot the induced interaction in Fig.~\ref{fig:Uq} for various fillings of the 2D system. In what follows, the mediated interaction for momentum transfers consistent with the marginal scattering processes [4, 9] play the key role in the development of various quantum phases of the system. At $\mu_f = 0$, $|U'(\pi)| > |U'(0)|$. However, as we increase $\mu_f$, $U'(0)/U'(\pi)$ becomes larger, and around $\mu_{f} \approx 1.8 t_f$, the magnitudes of $U'(0)$ and $U'(\pi)$ switch; this signals the appearance of different leading order instabilities, as will be shown below. With the original interaction $U_c$ included, the 1D system now has the total interaction $V(k_1, k_2, k_3)$ as
\begin{equation}
V(k_1,k_2,k_3) = U_c + U'(k_3-k_2),
\end{equation}
where the fourth momentum (not explicitly written above) is automatically determined by the momentum conservation. It follows that by varying the interactions, $U_c$ and $U_{cf}$, and the 2D filling $\mu_f$, we can control $U'(0)$ and $U'(\pi)$ to any negative value, while the overall strength of the interaction must be small enough for the perturbative scheme to be valid.

\section{Method}\label{sec:method}

\begin{figure}[!tbp]
\begin{center}
   \includegraphics[width = 0.9\columnwidth ]{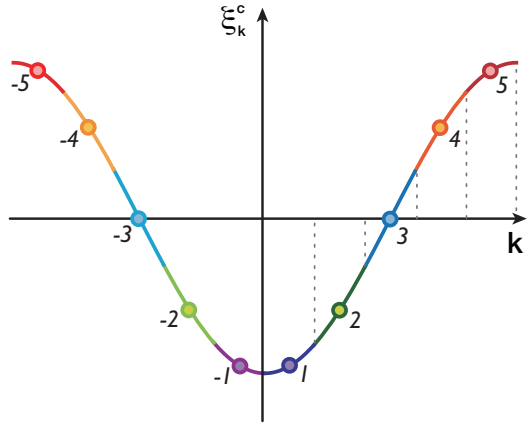}
   \caption{Patching scheme for the functional renormalization group.}
\label{fig:patch}
\end{center}
\end{figure}

Based on the effective interaction, phase diagrams are obtained by a fRG scheme. Here we briefly outline the standard $N$-patch scheme \cite{shankar1994, tam2006, tam2014, kopietz2010, platt2013}, which we employ in this work. We divide the Brillouin zone into $N_\text{patch}$ patches as shown in Fig.~\ref{fig:patch}, where the patch momenta $\{ \bar{k}_n \}$ are equally spaced. The interaction is now approximated as $V (k_1, k_2, k_3) \rightarrow   V_{n_1 n_2 n_3}$, where $n_i$ is the patch that $k_i$ belongs to. Naively the total number of coupling constants is $N_\text{patch}^3 = 27000$. However, we can reduce this number by using the symmetry of the Hamiltonian to 3735. The RG equation is obtained after integrating out the high-energy degrees of freedom around the ultraviolet cutoff $\Lambda$. By parametrizing the cutoff as $\Lambda(l) = \Lambda_0 e^{-l}$ with the initial value of the cutoff $\Lambda_0$, the coupling constants at lower energies are obtained by integrating the RG equations \cite{kopietz2010, platt2013}:
\begin{widetext}
\begin{equation}
\begin{split}
\frac{\partial  V_{n_1 n_2 n_3}}{\partial l} &= - \sum_{n} \dot{\Pi}^- (n, q_\text{pp}) \left( V_{n_1n_2n}V_{n_4 n_3 n} + V_{n_2 n_1 n}V_{n_3 n_4 n} \right)\\
 &+\sum_{n} \dot{\Pi}^+ (n, q_\text{fs})  \left( 2V_{n n_4 n_1}V_{n n_2 n_3} -V_{n_4 n n_1}V_{n n_2 n_3}- V_{n n_4 n_1}V_{n_2 n n_3}\right)\\
 &+\sum_{n} \dot{\Pi}^+ (n, -q_\text{fs})  \left( 2V_{n n_1 n_4}V_{n n_3 n_2} -V_{n_1 n n_4}V_{n n_3 n_2}- V_{n n_1 n_4}V_{n_3 n n_2}\right)\\
&- \sum_{n} \dot{\Pi}^+ (n, q_\text{ex}) V_{n_3 n n_1}V_{n_2 n n_4} - \sum_{n} \dot{\Pi}^+ (n, -q_\text{ex}) V_{n_1 n n_3}V_{n_4 n n_2},
\label{RG_Eq}
\end{split}
\end{equation}
\end{widetext}
where $q_\text{pp}=\bar{k}_{n_1}+\bar{k}_{n_2}$, $q_\text{fs}=\bar{k}_{n_3}-\bar{k}_{n_2}$, and $q_\text{ex}=\bar{k}_{n_1}-\bar{k}_{n_3}$. $\dot{\Pi}^{\pm} (n, q)$ is a differential of a bubble integral over frequency $\omega$ and momentum $k$ inside the $n$th patch,
\begin{equation}
\dot{\Pi}^{\pm} (n, q) = \pm \Lambda \int_{\omega} \int_{k \in n} \dot{G}(\omega, k) G[\pm \omega, \pm(k - q)],
\end{equation}
with $G(\omega, k) = \Theta(|\xi^c_k|- \Lambda)/(i\omega-\xi^c_k)$. Here the free propagator is used, since we ignore the self-energy correction along the RG flows.

A RG flow is started from an ultraviolet cutoff $\Lambda_0 = 2t_c$ and integrated until one of the coupling constants becomes $\sim 20t_c$ or $\Lambda = 10^{-6} t$. The former indicates an ordering instability, while the latter indicates no instability, {\it i.e.}, the Luttinger liquid fixed point. To determine the dominant instability, we extract the coupling constants at the Fermi level as
\begin{equation}
\begin{split}
g_1 &= V(L,R,L),\\
g_2 &= V(L,R,R),\\
g_3 &= V(L,L,R),
\end{split}
\end{equation}
where $L,R$ denote the left and right patches that the Fermi momentum is contained. This connects our treatment to the standard g-ology analysis, which usually assumes linear dispersion around the Fermi energy. In our analysis, the flows of these coupling constants are affected by the curvature of the band dispersion and by other marginal and irrelevant coupling constants away from the Fermi energy. As we discuss in the next section, these corrections are important to capture the emergence of a bond-order phase and subtle competition between pairing phases. We consider charge density-waves (CDW), spin density-waves (SDW), spin-singlet superconductivity (SS) and spin-triplet superconductivity (TS). When the system is at half-filling, bond charge/spin density-waves (BCDW/BSDW) are also possible. The ordering tendency of these phases are measured by \cite{menard2011}
\begin{equation}
\begin{split}
g_\text{CDW/BCDW} &= -2g_1 + g_2 (\mp g_3),\\
g_\text{SDW/BSDW} &= g_2 (\pm g_3),\\
g_\text{SS/TS} &= \mp g_1 - g_2,
\end{split}
\label{eq:g_MF}
\end{equation}
where the $g_3$ term is omitted for non-half-filling conditions. We identify the leading instability by the most diverging coupling constant in Eq.~\eqref{eq:g_MF}.

\section{Results}\label{sec:results}
\subsection{Phase diagrams}\label{sec:PD}
\begin{figure}[!tbp]
\begin{center}
   \includegraphics[width = \columnwidth ]{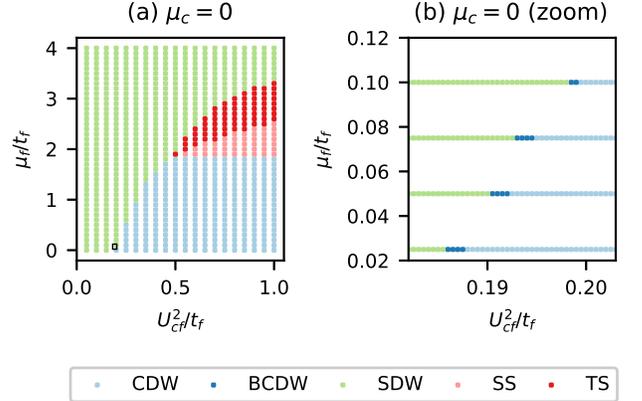}
   \caption{Phase diagrams for $\mu_c = 0$ (half-filling). The rectangular region in (a) is enlarged in (b) to zoom into the CDW-SDW boundary, showing the narrow BCDW phase.}
\label{fig:PD1}
\end{center}
\end{figure}
We start from the phase diagrams for half-filling, $\mu_c = 0$, in Fig.~\ref{fig:PD1}(a) at $U_c = t_c$. When the inter-species interaction $U_{cf}$ is weak or the 2D filling is very high, the total the interaction is dominated by $U_c$, which leads to the SDW instability. As $U_{cf}$ increases, the mediated attractive interaction becomes stronger. When the 2D fermions are half-filled $\mu_f = 0$ or close to it, the mediated interaction is dominated by the $q=\pi$ component, with $|U'(q=\pi)| > |U'(q=0)|$, as shown in Fig.~\ref{fig:Uq}, and CDW order becomes dominant. However, when the 2D filling deviates from half-filling significantly $\mu_f \gtrsim 1.8 t_f$, $U'(q=0)$ becomes the largest component
in the mediated attractive interaction (see Fig.\ref{fig:Uq}), which leads to spin-singlet pairing (SS). At $\mu_f \approx 1.8 t_f$, the mediated interaction is independent of momentum $q$, and thus gives attractive on-site interactions in real space. Hence at $U_{cf}^2/t_f \approx 0.55$, the original repulsive on-site Hubbard $U_c$ and the mediated contribution cancel, leading to a quadruple point where SDW, CDW, SS and TS meet (Fig.\ref{fig:PD1}). At this point, the system is non-interacting and behaves as a Luttinger liquid. Further increase of the 2D filling leads to spin-triplet pairing (TS) and then eventually the mediated interaction becomes too weak. We consider the TS region to be reminiscent of the $d$-wave pairing in a two-leg ladder with mediated interactions \cite{huang2013}. 

At half-filling, it is known that the extended Hubbard model exhibits BCDW at the boundary between CDW and SDW at weak-coupling regime \cite{nakamura2000, sengupta2002, tsuchiizu2002, tsuchiizu2004,sandvik2004,tam2006}. In Fig.~\ref{fig:PD1}(b), we plot the phase diagram near the CDW-SDW boundary to demonstrate the existence of the BCDW. While the BCDW region is rather narrow, our model correctly captures the known BCDW phase at the CDW-SDW boundary because of the radial patch scheme we employed for our fRG calculations. This region can be enlarged by increasing the original on-site interaction $U_c$, and the width of the BCDW is largest at $\mu_f = 0$, and becomes smaller as increasing $\mu_f$. This indicates that the BCDW can be enhanced by having larger values of $U'(\pi)$.

\begin{figure}[!tbp]
\begin{center}
   \includegraphics[width = \columnwidth ]{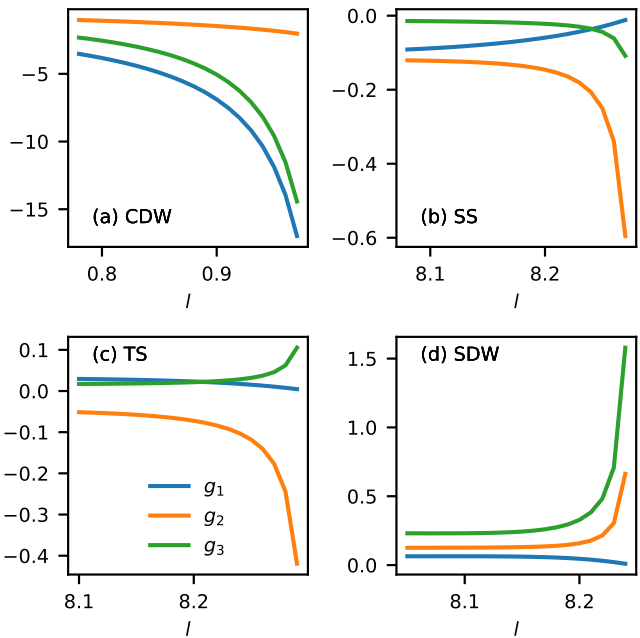}
   \caption{Flows of the coupling constants $g_{1,2,3}(l)$ for CDW (upper left), SS (upper right), TS (lower left), and SDW (lower right) regions at $U_{cf}^2/t_f = 0.8$. The values of $\mu_f/t_f$ are $0.0$, $2.1$, $2.7$, and $3.7$ respectively.}
\label{fig:flow}
\end{center}
\end{figure}

In order to further understand the origin of each region, we plot the flows of the coupling constants in Fig.~\ref{fig:flow}. In the CDW region, 
the large negative value of $g_1$ dominates. For $\mu_f \simeq 0$, the subdominant coupling is negative $g_3$ as in panel (a), while as we approach to the SS region, $g_2$ becomes subdominant. The growth of $g_2$ switches the dominant instability from the CDW to the SS. In the SS region [panel (b)] $g_2$ goes to $-\infty$, while $g_1$ and $g_3$ remain very small. Due to the small negative $g_1$, the SS instability is more dominant than the TS. In the TS region [panel (c)] $g_2$ still diverges to $-\infty$, while $g_3$ also grows, which signals the closeness to SDW. Here, due to the small positive $g_1$, the TS instability is more dominant than the SS. These indicate that the difference between the SS and TS instabilities is very subtle and determined by the marginally irrelevant $g_1$ coupling. The same behaviors are also seen in the extended Hubbard model in Ref.~\onlinecite{menard2011}. Finally, in the SDW region [panel (d)] both $g_2$ and $g_3$ diverge to $+\infty$ while $g_1$ is marginally irrelevant. Here, both the SDW and BCDW instabilities are possible, while the small positive $g_1$ makes the SDW more stable.

\begin{figure}[!tbp]
\begin{center}
   \includegraphics[width = \columnwidth ]{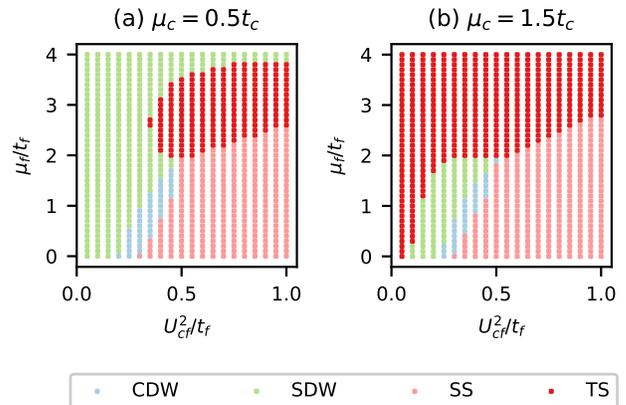}
   \caption{Phase diagrams for (a) $\mu_c = 0.5 t_c$ and (b) $\mu_c = 1.5 t_c$.}
\label{fig:PD2}
\end{center}
\end{figure}

Lastly, we show phase diagrams for $\mu_c \neq 0$ in Fig.~\ref{fig:PD2}. In general, $g_2$ becomes the most dominant term, and thus both SS and TS regions are enhanced. Compared to half-filling case, at $\mu_c = 0.5t_c$, the CDW region is greatly reduced, due to the absence of the Umklapp term, $g_3$, while the SDW region mostly remains. However, as we increase the filling to $\mu_c = 1.5t_c$, the SDW instability is replaced by the TS, and the density-waves only exist for relatively small $\mu_f$ and $U_{cf}$.

\section{Conclusions}\label{sec:conclusions}
In this work, we have studied the effect of mediated interactions on a Hubbard chain in contact with a two-dimensional non-interacting Fermi gas. After integrating out the fast non-interacting two-dimensional particles, we derive an effective mediated interaction among the one-dimensional fermions, whose momentum-dependent structure can be controlled by the filling of the mediating particles. With the mediated interaction and the original Hubbard interaction, we have obtained ground state phase diagrams of the one-dimensional fermions by the functional renormalization group. We find that the system can exhibit charge-/spin-/bond-density-waves and spin-singlet/-triplet superconducting instabilities by controlling the filling of the two-dimensional particles and the inter-species interaction. Effects of  trapping potentials, temperatures, and interactions among $f$-particles are interesting open problems.

\begin{acknowledgments}
J.O. acknowledges support by Georg H. Endress Foundation and by the state of Baden-W\"urttemberg through bwHPC. S.-W. T. acknowledges support from NSF under grant DMR-1411345. W.M.H acknowledges supports from the National Science Council in Taiwan through grant MOST 107-2112-M-005-008-MY3. 
\end{acknowledgments}

%

\end{document}